\documentclass[9pt, sigplan,nonacm]{acmart}
\AtBeginDocument{%
  }

\setcopyright{acmlicensed}
\copyrightyear{2018}
\acmYear{2018}
\acmDOI{XXXXXXX.XXXXXXX}
\acmConference[Conference acronym 'XX]{Make sure to enter the correct
  conference title from your rights confirmation email}{June 03--05,
  2018}{Woodstock, NY}
\acmISBN{978-1-4503-XXXX-X/2018/06}
\usepackage{algorithmic}
\usepackage{multirow}
\usepackage{textcomp}
\usepackage{tikz}
\usepackage{listings}
\usepackage{makecell}
\usepackage{float}
\usepackage{graphicx}
\usepackage{colortbl}
\usepackage[most]{tcolorbox}

\usepackage{booktabs}
\usepackage{amsmath}
\usepackage{caption}

\usepackage{indentfirst}
\usepackage{array, booktabs, multirow, cellspace, makecell, tabularx}
\usepackage{graphicx} 
\usepackage[table]{xcolor} 
\usepackage{flushend} 

\newcolumntype{Y}{>{\raggedright\arraybackslash}X} 
\newcolumntype{C}[1]{>{\centering\arraybackslash}m{#1}} 
\newcolumntype{L}[1]{>{\raggedright\arraybackslash}m{#1}} 
\tcbset{
  insightbox/.style={
    colback=gray!27,
    colframe=black,
    boxrule=0.7pt,
    arc=2pt,
    left=4pt,
    right=4pt,
    top=3pt,
    bottom=3pt,
    boxsep=0pt,
    fonttitle=\bfseries,
    before skip=6pt,
    after skip=6pt
  }
}

\usepackage[ruled,vlined]{algorithm2e}
\usepackage{amsmath} 
\usepackage{bm} 
\usepackage{pifont}
\usepackage[T1]{fontenc}
\usepackage[utf8]{inputenc}
\usepackage{subcaption}
\definecolor{GrayCodeBlock}{RGB}{241,241,241}
\definecolor{BlackText}{RGB}{0,0,0}
\definecolor{RedText}{RGB}{0,158,96}
\definecolor{BlueText}{RGB}{0,0,255}
\definecolor{DahongText}{RGB}{255,0,0}
\definecolor{RedTypename}{RGB}{182,86,17}
\definecolor{GreenString}{RGB}{96,172,57}
\definecolor{PurpleKeyword}{RGB}{184,84,212}
\definecolor{GrayComment}{RGB}{170,170,170}
\definecolor{GoldDocumentation}{RGB}{180,165,45}

\begin{document}



\title{AkiraRust: Re-thinking LLM-aided Rust Repair Using a Feedback-guided Thinking Switch}

\author{
    Renshuang Jiang$^{\dagger}$, 
    Yichong Wang$^{\dagger}$, 
    Pan Dong$^{\dagger}$, 
    Xiaoxiang Fang$^{\dagger}$, 
    Zhenling Duan$^{\dagger}$, \\
    Tinglue Wang$^{\P}$, 
    Yuchen Hu$^{\P}$, 
    Jie Yu$^{\dagger}$, 
    Zhe Jiang$^{\P}$ \\
     $^{\dagger}$National University of Defense Technology,  China, 
     $^{\P}$ Southeast University, China
}

\renewcommand{\shortauthors}{}

\begin{abstract}
   Eliminating undefined behaviors (UBs) in Rust programs requires a deep semantic understanding to enable accurate and reliable repair. 
   While existing studies have demonstrated the potential of LLMs to support Rust code analysis and repair, most frameworks remain constrained by inflexible templates or lack grounding in executable semantics, resulting in limited contextual awareness and semantic incorrectness. 
   Here, we present AkiraRust, an LLM-driven repair and verification framework that incorporates a finite-state machine to dynamically adapt its detection and repair flow to runtime semantic conditions. AkiraRust introduces a dual-mode reasoning strategy that coordinates fast and slow thinking across multiple agents. 
   Each agent is mapped to an FSM state, and a waveform-driven transition controller manages state switching, rollback decisions, and semantic check pointing, enabling context-aware and runtime-adaptive repair.   
   Experimental results show that AkiraRust achieves about 92\% semantic correctness and delivers a 2.2× average speedup compared to SOTA.
\end{abstract}


\newcommand{\zhenote}[1]{\textbf{\textcolor{blue}{Zhe: #1}}}
\newcommand{\shuangote}[1]{\textbf{\textcolor{magenta}{Shuang: #1}}}
\newcommand{\XXnote}[1]{\textbf{\textcolor{orange}{Xiang: #1}}}
\newcommand{\DPnote}[1]{\textbf{\textcolor{green}{DP: #1}}}
\maketitle

\vspace{-1.0em}
\section{Introduction}
\label{introduction}
Rust language~\cite{matsakis2014rust} is designed for system-level programming with strong guarantees of memory safety, and has been widely adopted in critical domains such as operating systems and browsers \\~\cite{levy2015ownership,levy2017case,anderson2016engineering}. 
However, for the flexibility and compatibility with low-level operations, Unsafe Rust allows developers to bypass the compiler’s safety checks, reintroducing potential undefined behaviors (UBs). 
Rust’s safety mechanisms, while ensuring robustness, complicate the semantic understanding and repair of UBs, even minor changes must respect strict compiler checks.  
Studies show that repairing and verifying Rust’s safety properties have become critical aspects of software development, accounting for over 40\% of total debugging costs~\cite{astrauskas2019leveraging}. 
Therefore, how to achieve automated, reliable and semantically correctness repair of Rust has become a challenge. 

\begin{figure}[t]
\centerline{\includegraphics[width=0.47\textwidth]{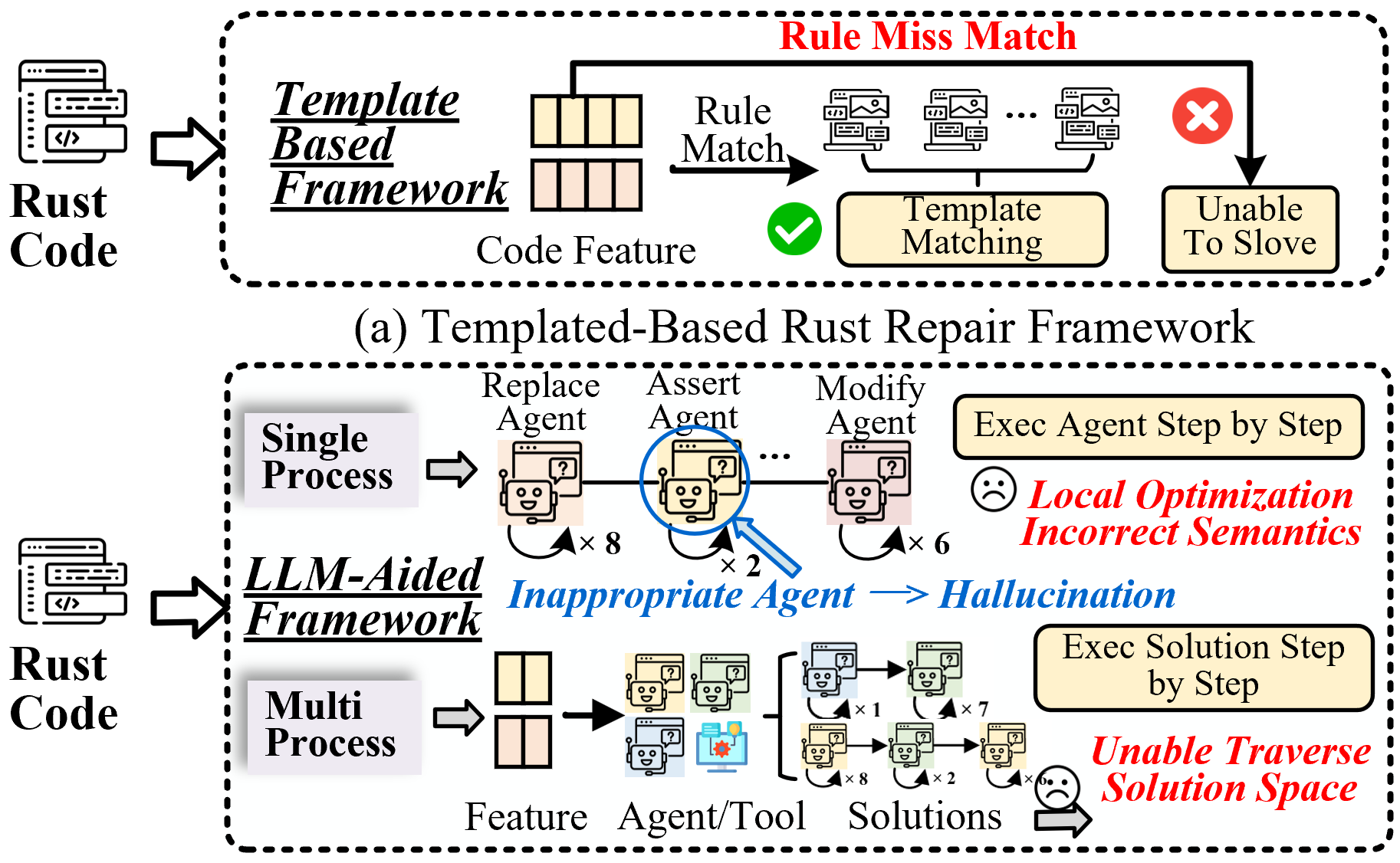}}
\vspace{-1.2em}
\caption{Existing Rust Repair Frameworks. 
(a) Template-Based: Rely on pattern matching; out-of-scope patterns lead to failed repairs.
(b) LLM-Aided: Single-process with a fixed agent pipeline, while multi-process dispatch agents by heuristic features. Unable to traverse solution space and semantically inaccurate.
}
\label{high-level}
\vspace{-1.0em}
\end{figure}

\noindent
\textbf{Existing Works.}
Traditional static and dynamic methods, including formal, fuzzing, and symbolic execution (e.g., Thetis-Lathe~\cite{jiang2025thetis}, RULF~\cite{jiang2021rulf}, and SafeNet~\cite{dong2024safenet}), remain template-based and rely on rule-matching logic. 
When errors extend predefined patterns or involve complex semantics, these fail due to the lack of semantic understanding (Fig.~\ref{high-level}a).  
LLM-aided frameworks show promising capabilities in Rust repair~\cite{jiang2025unlocking,wu2023rustgen,deligiannis2025rustassistant}, leveraging deeper semantic understanding and contextual reasoning ability. 
However, existing frameworks treat repair as an offline single-process, where agents and their order are fixed before execution, even though the actual repair runs online (Fig.~\ref{high-level}b), e.g., RustAssistant~\cite{deligiannis2025rustassistant} and RustGen~\cite{wu2023rustgen}. 
While leveraging LLM capabilities allows them to handle a broader range of static template-based errors, inappropriate agent invocation can still lead to semantic violations.  
For instance, borrowing-related Unsafe Rust errors cause template-based schemes (e.g., “assert–modify’’) to repeatedly insert assertions, leading to deviation and hallucination; switching to modification agents often traps the model in local code adjustments without performing the required semantic fix. 
Multi-process frameworks (e.g., RustBrain~\cite{jiang2025unlocking}, RepairAgent~\cite{bouzenia2024repairagent}) attempted to consider more Rust features to provide offline multi-process candidates, but they only based on initial code features (Fig.~\ref{high-level} (b)), also suffering from semantic incorrectness. 

\noindent
\textbf{Challenges.} 
When we attempted to develop a framework to address Rust’s security and semantic correctness issues, we found that these frameworks commonly achieve high pass rates under security detection tools (like Miri~\cite{Miri_dataset}, Kani~\cite{vanhattum2022verifying}), but their repairs frequently deviate from the intended semantics, revealing a lack of dynamic adaptability in repair processes. 
From the reasoning perspective, current static and LLM-based frameworks either operate within inflexible template constraints or lack grounding in executable semantics. 
Due to Rust’s strong type system and ownership semantics, the selection and coordination of repair agents significantly influence the reasoning path and repair accuracy (see Sec.~\ref{observation1}), restricted template schemes frequently cause reasoning deadlocks or oscillations, preventing the model from discovering valid repair strategies in dynamic contexts. 
From the runtime perspective, repair processes rarely incorporate runtime signals such as execution traces, borrow-state transitions, or verification, resulting in limited semantic awareness and fragile validation of generated fixes. 
All of these limitations hinder the dynamic adaptability and semantic correctness required for accurate and efficient Rust repair. 

\noindent
\textbf{Contributions.}
Here, we present AkiraRust, an LLM-driven Rust repair framework that integrates finite state machine (FSM) to dynamically adapt its repair processes to runtime semantic variations. 
By leveraging FSM, 
(i) the states of AkiraRust are composed of a dual-mode multi-agents library, mapping multiple agents and thinking modes (fast and slow thinking) to distinct states, thereby enabling dynamic strategy selection and improving repair accuracy and efficiency. 
(ii) the state transition mechanism controls the selection of agent states and the construction of reasoning sequences, combining runtime monitoring signals to track the repair process and validate semantic correctness, while guiding the LLM to self-correct or adjust hallucinated repair suggestions. 
As a demonstration on the Miri dataset~\cite{Miri_dataset}, AkiraRust achieves about 92\% semantic correctness and delivers a 2× speedup over SOTA.



\section{Background and Motivation}
\label{motivation}


\subsection{Agent Selection for Adaptive Repair}
\label{observation1}
Due to Rust’s strict semantics, the repair process must maintain fine-grained semantic correctness.  
Unsafe Rust, though limited in pattern variety~\cite{astrauskas2020programmers,zhang2023dual, jiang2023thetis}, is highly sensitive, requiring precise reasoning over memory boundaries and cross-module.  
For LLM-aided repair, different agents are suited to different defect types. 
We use RustBrain~\cite{jiang2025unlocking}, a SOTA repair tool, to generate ten solutions and record multiple repair configurations with different agent combinations and execution orders to evaluate how these affect semantic correctness (Tab.~\ref{insight1}). 
Here, $N$ is the total number of repair rounds, $q_{\text{total}}$ the overall semantic accuracy, and $q_{\text{agt}}$ the effectiveness of each RustBrain agent: $q_{\text{assert}}$ for assertion insertion, $q_{\text{modify}}$ for code modification, $q_{\text{replace}}$ for Safe~Rust replacement, and $q_{\text{knowledge}}$ for RAG querying. 
Higher $q_{\text{agt}}$ indicates stronger agent specificity.

\begin{table}[hbpt]
\centering
\vspace{-0.5em}
\caption{Effectiveness of Agents in Unsafe Rust Cases.} 
\vspace{-1.0em}
\setlength{\tabcolsep}{1.7pt} 
\footnotesize
\begin{tabular}{c l c c c c c c}
\toprule
\textbf{Index} & \textbf{Case Feature} & \textbf{$N$} & \textbf{$q_{total}$} & \textbf{$q_{assert}$} & \textbf{$q_{modify}$} & \textbf{$q_{replace}$} & \textbf{$q_{knowledge}$} \\
\midrule
1 & size mismatch & 14 & 0.50 & 0.286 & 0.286 & \cellcolor{red!23}{0.429} &  0.286 \\
2 & access violation & 14 & 0.50 & \cellcolor{red!23}{0.429} & 0.286 & \cellcolor{pink!23}{0.357} & 0.286 \\
3 & write access & 13 & 0.692 & 0.615 & \cellcolor{cyan!23}{0.385} & 0.615 &  0.615 \\
4 & retag write & 12 & 0.75 & \cellcolor{red!23}{0.75} & 0.417 & 0.583 & 0.583 \\
5 & dangling pointer & 11 & 0.909 & 0.545 & 0.636 & \cellcolor{red!23}{0.727} & 0.273 \\
6 & atomic access & 10 & 1.00 & 0.600 & 0.600 & \cellcolor{cyan!23}{0.200} &  \cellcolor{cyan!23}{0.250} \\
\bottomrule
\vspace{-2em}
\end{tabular}
\label{insight1}
\end{table}

In Tab.~\ref{insight1}, cases 1, 2, 4, and 5 show that a properly selected single agent nearly matches $q_{total}$, indicating that an appropriately selected single-agent already captures the core structure, and additional agents contribute little beyond redundant hypotheses. 
This is because unsafe Rust errors arise from a finite set of structured operations, e.g., size mismatches require semantic-preserving replacements, while access violations demand safety assertions, making certain agents inherently better suited to specific defect types. 
Cases 3 and 6 show similar success rates across agents, as these defects require multi-agents coordination, but inappropriate actions (e.g., $q_{modify}$ in case 3) can cause semantic inconsistencies. 

\begin{tcolorbox}[insightbox]
\small
\textbf{Insight 1:} Aligning agent with defect features is crucial, as appropriate coordination markedly enhances repair and semantic accuracy.  
\end{tcolorbox} 

\subsection{Reasoning Modes for Adaptive Repair}
\label{observation2} 
While LLMs possess strong capabilities in language understanding and pattern recognition, their effectiveness in repair tasks is highly dependent on the underlying reasoning mode. 
LLMs exhibit two distinct modes of thinking~\cite{kahneman2011fast,de2023advancing,qu2025survey}: fast thinking and slow thinking. 
Fast thinking is quick and intuitive, producing candidate fixes with minimal computation. 
Slow thinking, in contrast, is more careful and step-by-step. 
It examines the code structure in more detail, checks ownership and lifetime rules more thoroughly, and reasons across multiple steps before proposing a repair. 
By selecting 100 defect samples from the official Rust library and open-source projects~\cite{Rust_libaray}, we categorized Rust code by repair complexity levels (1–5) based on the LOC~\cite{loc} and experienced engineers' classificationclassify. 
We then compared the semantic accuracy and efficiency of fast and slow thinking across these complexity levels, where efficiency is measured by the total time required to complete a repair. 

\begin{figure}[htbp]
\centerline{\includegraphics[width=0.38\textwidth, trim=0cm 0.02cm 0cm 0cm, clip]{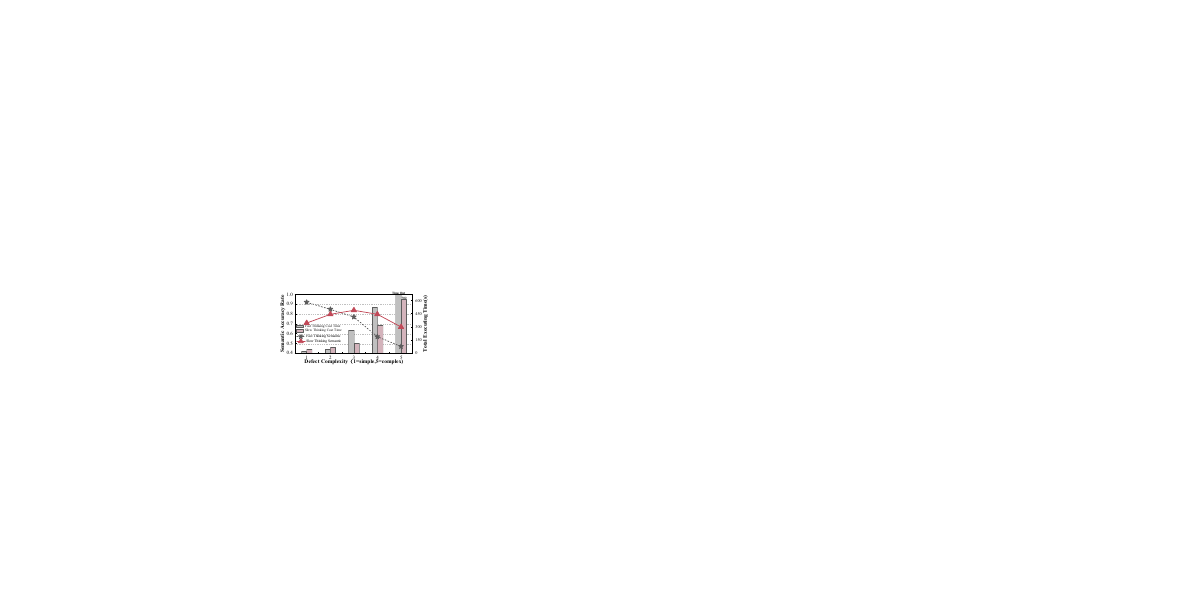}}
\vspace{-1.0em}
\caption{Fast vs. Slow Thinking in Rust Repair.}
\vspace{-1em}
\label{obs2}
\end{figure}

In Fig.~\ref{obs2}, for simple defects, fast thinking achieves a repair success rate of 92\% and completes repairs with minimal iterations, demonstrating higher efficiency than slow thinking, which tends to over-reason in certain cases, introducing redundant logic or hallucinated code. As defect complexity increases, however, slow thinking enhances accuracy and interpretability, surpassing fast thinking for complex, context-dependent repairs. In these cases, fast thinking requires multiple iterations, leading to more than a 3× overhead.

\begin{tcolorbox}[insightbox]
\small
\textbf{Insight 2:} Effective repair requires switching between fast and slow thinking based on Rust code features to balance efficiency and accuracy. 
\end{tcolorbox} 

\subsection{Hallucination Control for Adaptive Repair} 
\label{observation3}
LLMs may produce incorrect or semantically inconsistent reasoning during the repair process, known as hallucination~\cite{ji2023survey,ji2023towards,yao2023llm}. 
Each reasoning step can be regarded as a decision under uncertainty, we design a code-state evaluation method, using UBs number, semantic consistency, and structural soundness, to measure the code incorrectness (see details in Sec.~\ref{adptive_rollback}). 
We then define the hallucination score as the deviation between the maximum code incorrectness score and the initial code. 
Since temperature directly affects the stochasticity of model reasoning, we construct three comparative pipelines: (i) low temperature w/o rollback\footnote{Rollback: the framework reverses back to the initial code when an anomaly is detected.}; (ii) high temperature w/o rollback; (iii) high temperature w/ rollback, to analyze how hallucination control influences repair quality. 

\begin{figure}[htbp]
\vspace{-1em}
\centerline{\includegraphics[width=0.45\textwidth, trim=0cm 0.25cm 0cm 0cm, clip]{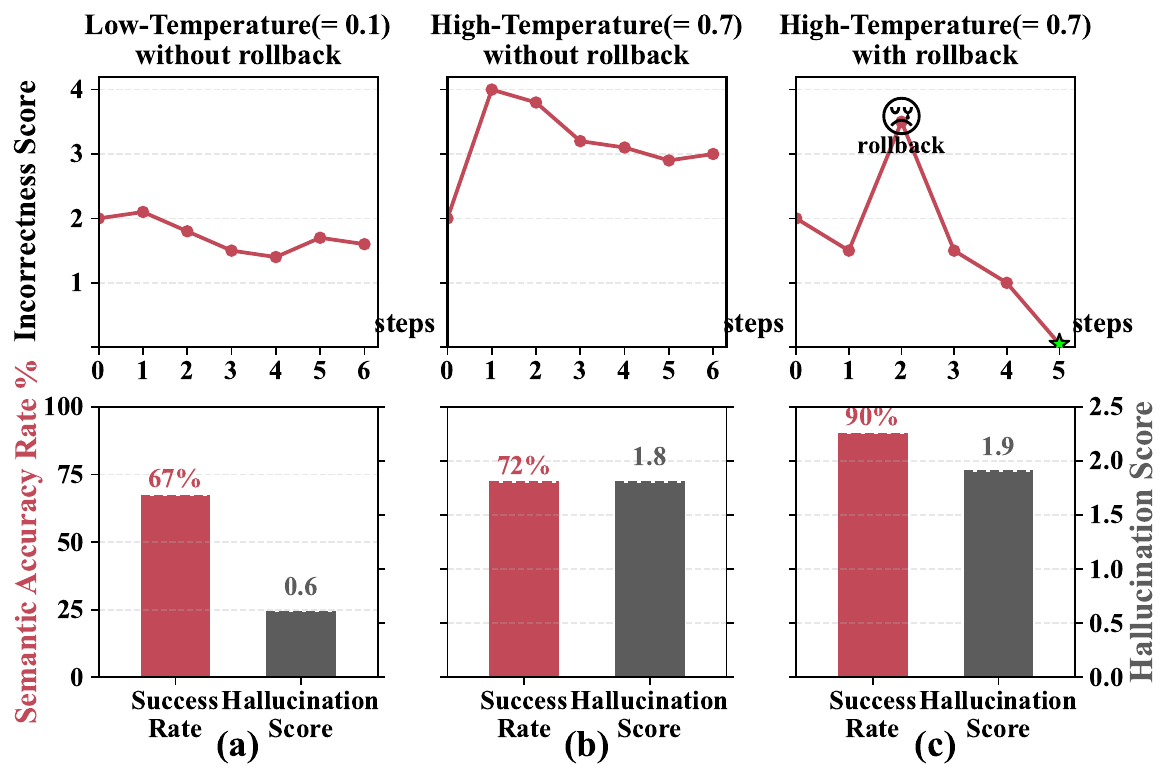}}
\vspace{-1.0em}
\caption{Temperature and Rollback Effects on Rust Repair.}
\vspace{-1em}
\label{obs3}
\end{figure}
 
In Fig.~\ref{obs3}, the line plot shows the evolution of incorrectness scores across repair steps, while the bar plot reports the final repair success rate and average hallucination deviation.
Results indicate that high-temperature reasoning is more divergent, yielding both higher hallucination levels and higher repair success rate (Fig.~\ref{obs3} (a) and (b)). Introducing rollback under high temperature, effectively suppresses hallucination, improves the accuracy of repair semantics and reduces the number of agent invocations (Fig.~\ref{obs3} (b) and (c)). 

\begin{tcolorbox}[insightbox]
\small
\textbf{Insight 3:} Hallucinations inherently have exploratory potential, and the point is to harness and control them via verification mechanisms, enabling improved repair semantic accuracy and efficiency.  
\end{tcolorbox}

\section{AkiraRust: At a Glance} 
\begin{figure}[htbp]
\vspace{-0.8em}
\centerline{\includegraphics[width=0.5\textwidth]{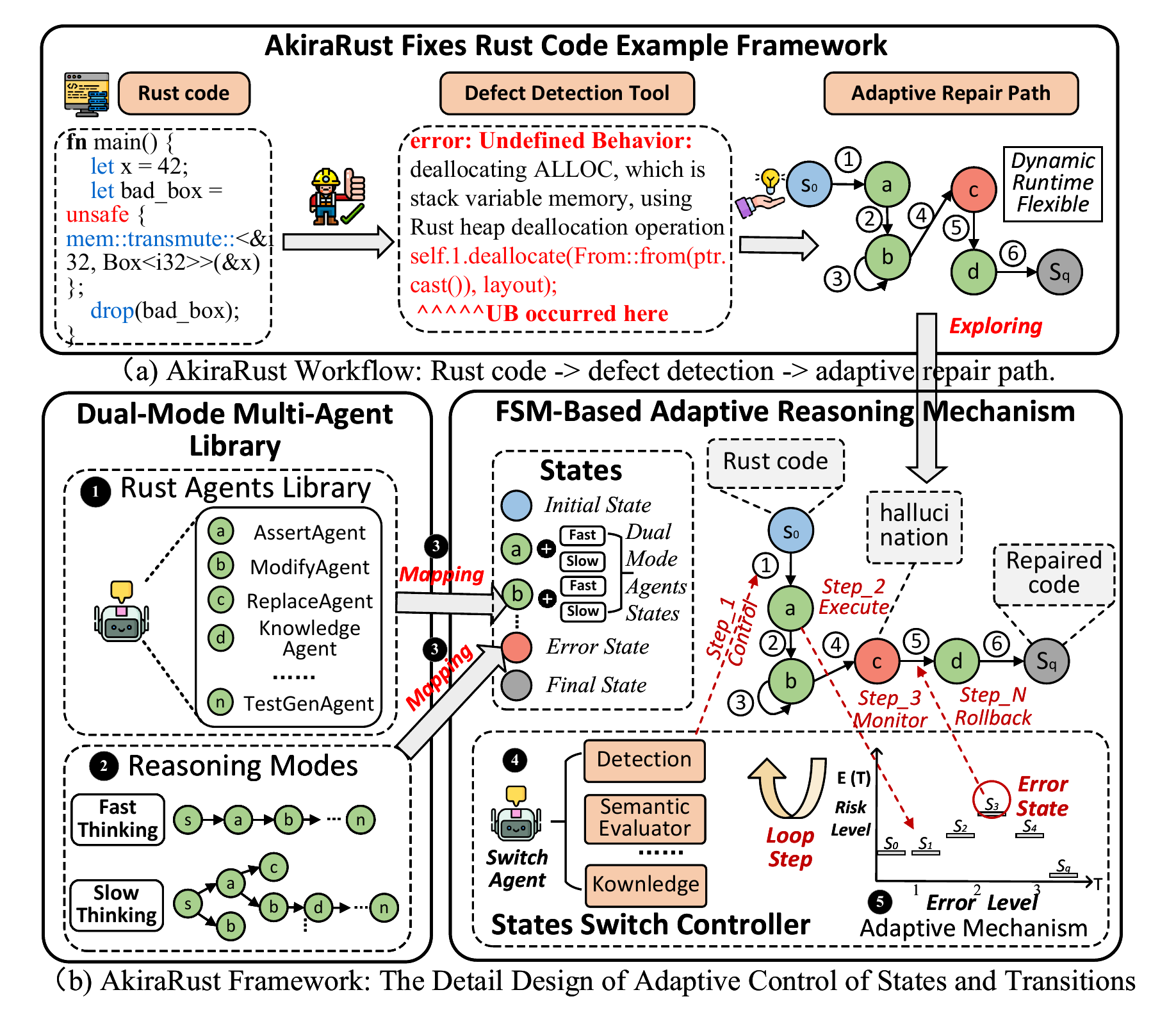}}
\vspace{-1em}
\caption{AkiraRust Framework: 
(a) illustrates the AkiraRust workflow. 
(b) presents the detailed design of AkiraRust. 
\ding{182} and \ding{183} integrate a dual-mode multi-agent library, mapping to the FSM states in \ding{184}, while 
\ding{185} and \ding{186} guide state transitions.
}
\vspace{-1em}
\label{framework}
\end{figure}

AkiraRust systematically addresses the challenges by FSM-based runtime adaptive reasoning and repair orchestration, enabling the framework to determine \textbf{which} agents or tools to select, \textbf{how} reasoning mode to adopt, and \textbf{what} actions to halt or rollback during the repair process. 
Overall, according to Rust’s semantic features, AkiraRust is composed of multiple LLM-based agents that operate under fast and slow thinking modes, and generate candidate repairs by diverse strategies. 
Then, AkiraRust constructs an FSM that formally describes each state with agents and thinking modes, and defines transition conditions for selection states. 
By performing runtime analysis of code, AkiraRust dynamically constructs waveform evolution, which is a code-incorrectness scores curve that reflects how the repair quality changes over time, and uses it to guide rollback decisions and adjust the reasoning depth.  
Fig.~\ref{framework}(a) presentshows the workflow of AkiraRust framework, showing the process from Rust code by UB detection to the adaptive repair path. 

\vspace{-0.5em}

Detailed design in Fig.~\ref{framework}(b), existinging Rust researches show that~\cite{jiang2025unlocking,dong2024safenet,qin2020understanding} Rust repair method can be categorized into three types:  (i) replacements with Safe Rust equivalents, (ii) assertion-based prevention, and (iii) semantic-level modifications.  
Based on these, we developed a multi-agent repair library, where each agent invokes specialized repair templates tailored to different unsafe patterns and error categories to achieve semantically consistent repairs (\ding{182} in Fig.~\ref{framework}). 
From Insight 1, the multi-agent repair library provides targeted repair strategies for type-related defects, e.g., assertion agents are used for type-mismatch errors, and modification agents handle value-correction errors, thereby offering more appropriate agent choices, reducing the overhead caused by random agent selection, and improving repair accuracy. 
Furthermore, from Insight 2, during code generation, the LLM’s fast and slow thinking modes significantly influence agents’ reasoning depth and repair path selection. 
For the same agent, different thinking modes may yield distinct repair strategies.
To better accommodate this cognitive diversity, we construct a dual-mode multi-agent library (\ding{183} in Fig.~\ref{framework}), which modularized fast-thinking agents (heuristic and lightweight repair) and slow-thinking agents (deep semantic repair). 
This design enables adaptive invocation of appropriate agents across reasoning stages, achieving more flexible and efficient repair.

To address the limitation of inflexible template constraints and lacking executable semantics (Sec.~\ref{introduction}), 
We construct an FSM-based repair framework that allows each agent to adaptively explore repair paths at runtime and gradually correct them under guidance from code-context semantic feedback.  
The dual-mode multi-agent library maps agents and modes to the corresponding states of the FSM to support adaptive selection (\ding{184} in Fig.~\ref{framework}). 
The FSM’s state-transition mechanism captures the current repair context, including the UBs number, semantic correctness, and code structural, and determines the next agent invocation and reasoning mode (\ding{185} in Fig.~\ref{framework}). 
However, each transition introduces a semantic checkpoint, a significant evaluation overhead is inevitably incurred. 
From Insight 3, we design a self-adaptive correction mechanism based on hallucination-aware evaluation to enhance state transitions conditions (\ding{186} in Fig.~\ref{framework}). 
It scores the incorrectness of each generated code, treating semantic hallucinations as deviations from expected behavior; whenever the incorrectness exceeds a defined threshold, it triggers rollback or semantic evaluation to prevent error propagation and maintain repair stability. 

Overall, AkiraRust abstracts the repair process as a form of human-like reasoning, where each state transition reflects a cognitive shift between exploration and reflection. 
By embedding adaptation into the repair path, it breaks the rigidity of static processes and enables the exploration of a larger repair space. 
Besides, compared to running all agents at every step, it avoids combinatorial search explosion and excessive overhead. 
AkiraRust thus enables context-aware and semantically consistent repair of Rust UBs, enhancing the correctness and safety of the repaired code. 

\vspace{-0.5em}
\section{Implementation of AkiraRust}
\subsection{FSM as the Backbone of Repair}
\label{fsm}
An FSM is composed of a finite set of states, a set of input symbols, transition functions, and output mappings that describe the evolution of the system over time~\cite{brand1983communicating, linz2022introduction}. 
In our setting, each state represents a distinct repair agent or thinking mode, reflecting different cognition or operations during the repair process. 
State transition conditions are determined by runtime semantic feedback signals, including repair success, semantic correctness, and hallucination likelihoods. 
Therefore, the FSM can be formally defined as a six-tuple $FSM=(\Sigma,S,s_{0},F,\delta,\lambda)$, where: 
\begin{itemize}
    \item \textbf{Input}:  
    $\Sigma = \{ r \mid \text{Rust Code} \}$,  
    representing the set of input programs or repair candidates evaluated during the process.
    \item \textbf{State Set}: Each state corresponds to a specific phase: 
    $S = \{ q_0, q_{assert}, q_{modify}, q_{replace}, q_{knowledge}, q_{rollback}, q_{Err}, q_F \}$. 
    $q_0$ and $q_F$ are the initial and final states, $q_{Err}$ terminates repair afteron repeated failures, and the others are defined in Sec.~\ref{observation1}. 
    \item \textbf{Initial State}:  
    $s_0 = q_0$
    \item \textbf{Final State Set}:  
    $F = \{ q_F, q_{Err} \}$
    \item \textbf{State Transition Function}:  
    $\delta: S \times \Sigma \rightarrow S$, defines FSM transitions driven by code-context.
    \item \textbf{Output}:  
    $\lambda: S\rightarrow \Sigma$, refers Rust repair code.
\end{itemize} 

For a Rust program detected with UBs by static or dynamic tools (e.g., Miri~\cite{Miri_dataset}), AkiraRust initializes it as the FSM’s initial state $q_0$, with input symbols $\Sigma$ determined by the error type and semantic context. 
Under feedback-driven transition conditions, the FSM dynamically invokes appropriate repair agents to generate repair code.
When the defect is successfully fixed and passes compilation and semantics checks, the FSM reaches the final state $q_F$; otherwise, after exceeding the maximum transition limit, it enters the error state $q_{Err}$, indicating repair failure. 
Next, we present the design of the dual-mode multi-agent library is detailed, including its mapping to the FSM state set and the corresponding transition mechanism.

\subsubsection{\textbf{Dual-Mode Multi-Agents State}}
\label{multi-agent} 
Existing Rust automated frameworks rely on fixed templates and rigid orchestration, resulting in inflexible repair processes. 
Compared to C language unsafe operations~\cite{emre2021translating,ling2022rust}, the finiteness of Unsafe Rust provides a well-defined opportunity to systematically categorize and manage repair strategies.  
Firstly, the repair patterns are limited and can be classified into three types: replacement with Safe Rust, assertion-based prevention, and semantic modification~\cite{jiang2025unlocking,dong2024safenet}. 
Then, LLM thinking modes can be divided into two typical approaches: fast thinking performs rapid repairs based on heuristics and contextual matching, whereas slow thinking conducts deep semantic reasoning and contextual consistency analysis. 
We abstract the above repair patterns and reasoning modes into a set of orthogonal basis vectors to construct a dual-mode multi-agent library.   

\begin{figure}[t]
\centerline{\includegraphics[width=0.46\textwidth]{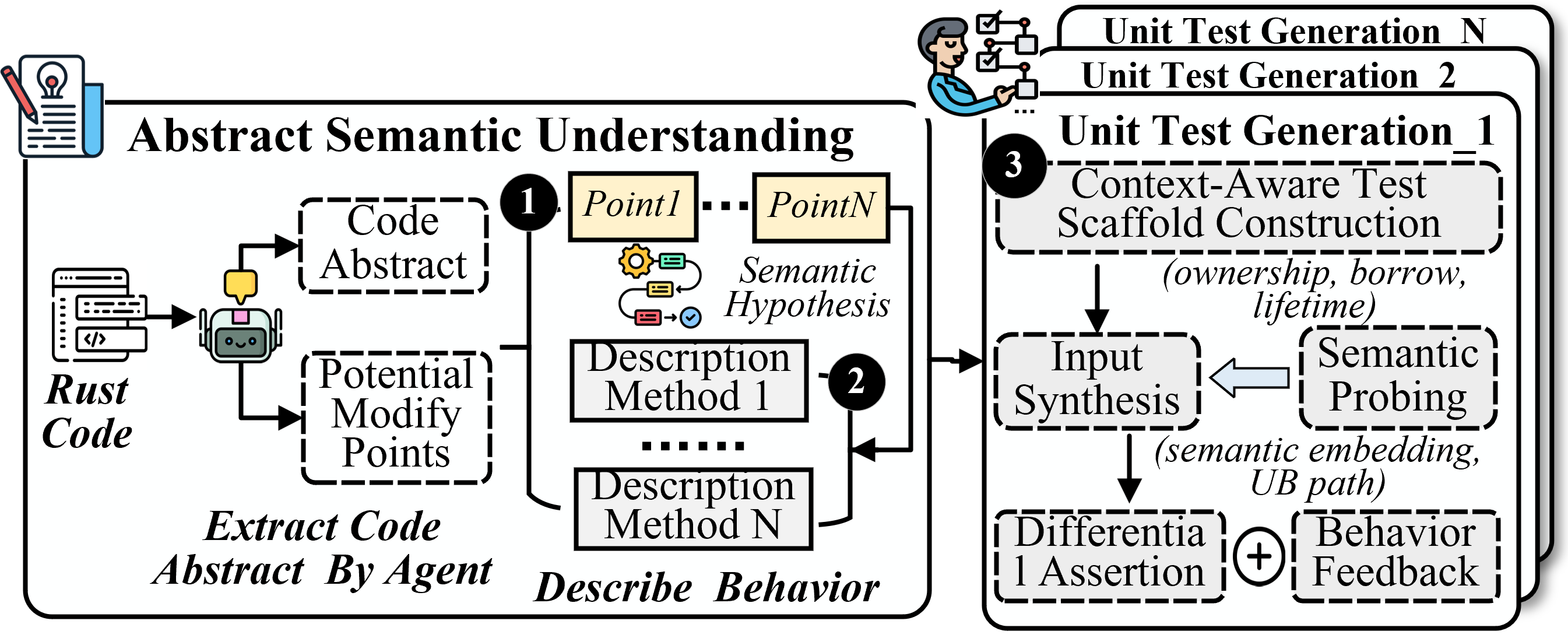}}
\vspace{-1.0em}
\caption{TestGenAgent: User-Intent–Guided Test Agent. }
\vspace{-1.0em}
\label{unit_test}
\end{figure}

To fully leverage existing capabilities and reduce redundant implementation, existing Rust repair tools and specialized agents are integrated into the framework, including assertion, modification, replacement, and knowledge agents.  
Besides, we introduce evaluation agents to perform semantic validation.  
Rust’s memory safety mechanism makes semantic correctness challenging, even experienced developers spend over 15\% of their time writing unit tests~\cite{zhu2022learning,cheng2025rug}.   
While compiler tools detect memory and type errors, they do not capture higher-level semantic intent or behavioral correctness (e.g., unsafe blocks that compile but violate aliasing).  
Existing LLM-aided semantic validation methods, like CodeJudge~\cite{tong2024codejudge} rely on prompt engineering with other LLMs to judge semantic; 
test generation methods (e.g, RUG~\cite{cheng2025rug}, Rust-twins~\cite{yang2024rust}) focus on syntactic corrections or surface-level consistency, neglecting the underlying semantic. 
However, the correct repair ultimately depends on the developer’s intention. 
For instance, in cases of misaligned \textit{alloc} and \textit{dealloc}, the goal is to ensure memory alignment, but which variable to adjust and how to align it requires understanding the intended behavior. 
Therefore, we propose a \textit{TestGenAgent} that leverages Rust’s strong type system and the built-in testing pipeline (``cargo test'') to perform intent-guided test generation and semantic validation. 
Given a repaired candidate, TestGenAgent first summarizes key code features and infers potential modification points (\ding{182} in Fig.~\ref{unit_test}), such as mismatched allocation–deallocation pairs or alignment-sensitive memory operations. It then analyzes these points to derive intent-aligned semantic constraints and generate candidate behavioral conditions (\ding{183} in Fig.~\ref{unit_test}).
Based on these constraints, the agent produces repair-aware semantic variants and synthesizes unit tests that exercise the inferred behaviors; if at least one variant passes all generated tests, the repair is considered acceptable (\ding{184} in Fig.~\ref{unit_test}). 
This process enables exploring a wider range of repair variations, improving test coverage while ensuring the repaired code’s semantics remain within the intended behavioral bounds. 

\noindent
\textbf{Accuracy.} We discuss the false negative/positive~\cite{False_Positives} of \textit{TestGenAgent} to analyze its accuracy in semantic validation. 
The results show semantic evaluation achieves 85\% accuracy: 10\% false negative, where abnormal behaviors that may violate Rust’s safety or intended semantics are mistakenly classified as normal, potentially compromising program correctness; and 5\% false positive, where normal behaviors are misclassified as abnormal, leading to extra semantic checks or rollbacks and increased overhead (Tab.~\ref{flase_pos_neg}). 
\vspace{-1.0em}
\begin{table}[h!]
\centering
\caption{Semantic evaluation results of \textit{TestGenAgent}}
\vspace{-1.0em}
\small
\label{flase_pos_neg}
\begin{tabular}{l|c|c}
\toprule
 & \textbf{Predicted Abnormal} & \textbf{Predicted Normal} \\
\midrule
\textbf{Actual Abnormal} &  True Positive: 0.78 & \cellcolor{red!23} False Negative: 0.10 \\
\textbf{Actual Normal} & \cellcolor{cyan!23} False Positive: 0.05 & True Negative: 0.07 \\
\bottomrule
\end{tabular}
\end{table}
\vspace{-1em}

Overall, this dual-mode multi-agent library efficiently reuses existing agents and enables flexible composition of repair strategies, allowing agents to adapt adaptively based on code-context. 
These agents are further abstracted as distinct states within the FSM, with each state corresponding to a specific repair intention.

\subsubsection{\textbf{Transformation and Coordination among States}}
\label{transition} 
To capture the dynamic evolution of repair behaviors and enable coordinated reasoning across multiple agents, AkiraRust introduces a state transformation mechanism, which uses runtime indicators to represent the evolving repair context and adaptively select the most suitable next repair agent and modes from library states. 
Formally, the state transformation is expressed as:
\begin{equation*}
    \delta: (S_{t}, \alpha_{t}) \longmapsto (S_{t+1}, A_{t+1})
\end{equation*}
where $S_{t}$ is the current FSM state and $\Sigma_{t}$ is the vector of runtime signals, the next state $S_{t+1}$ and the agent action output $A_{t+1}$. 


The transformation mechanism operates in two complementary stages:  
(i) the front-end performs a fast, local evaluation of the current Rust code state, which a lightweight check derived directly from runtime signals, by inspecting the runtime signal vector $\alpha_t$, including UB counts, types, and semantic context from \textit{TestGenAgent}.
 When the UB count is zero, the current state is considered a ``stable candidate'' for subsequent verification. 
Otherwise, the framework uses the LLM to extract potential UB keyword tags to reflect the current semantic context. 
(ii) the back-end builds a \textit{FixAgent} that selects appropriate $S_{t+1}$ (agent and mode) based on $\alpha_t$. 
Given Rust’s strong typing and ownership rules, AkiraRust maintains a dynamic knowledge base that accumulates prior repair experience, enabling the \textit{FixAgent} to more accurately choose the next state $S_{t+1}$ and generate the corresponding action $A_{t+1}$. 
Overall, this transformation guides adaptive adjustment of repair path, achieving stepwise agent/mode selection and context-aware repair orchestration.

\subsection{Self-Adaptive Correction Mechanism}
\label{adptive_rollback} 
We analyze the overhead of AkiraRust within the FSM framework: 
(i) semantic evaluation is required for each intermediate Rust code, and (ii) hallucinations (reflected in the final code state) trigger direct rollback to the initial state. 
From Insight 3, hallucination depends not only on the performance of final states but also on the overall judgment across the entire repair process, direct rollback neglecting intermediate repair progress may increase semantic evaluation overhead.    
Therefore, AkiraRust implements a self-adaptive correction mechanism driven by waveform feedback to strength the transformation mechanism. 
It continuously monitors the runtime dynamics of the repair state and determines optimal evaluation or rollback point, reducing semantic validation and rollback overhead. 
\vspace{-2.0em}
\begin{figure}[h]
\centerline{\includegraphics[width=0.47\textwidth]{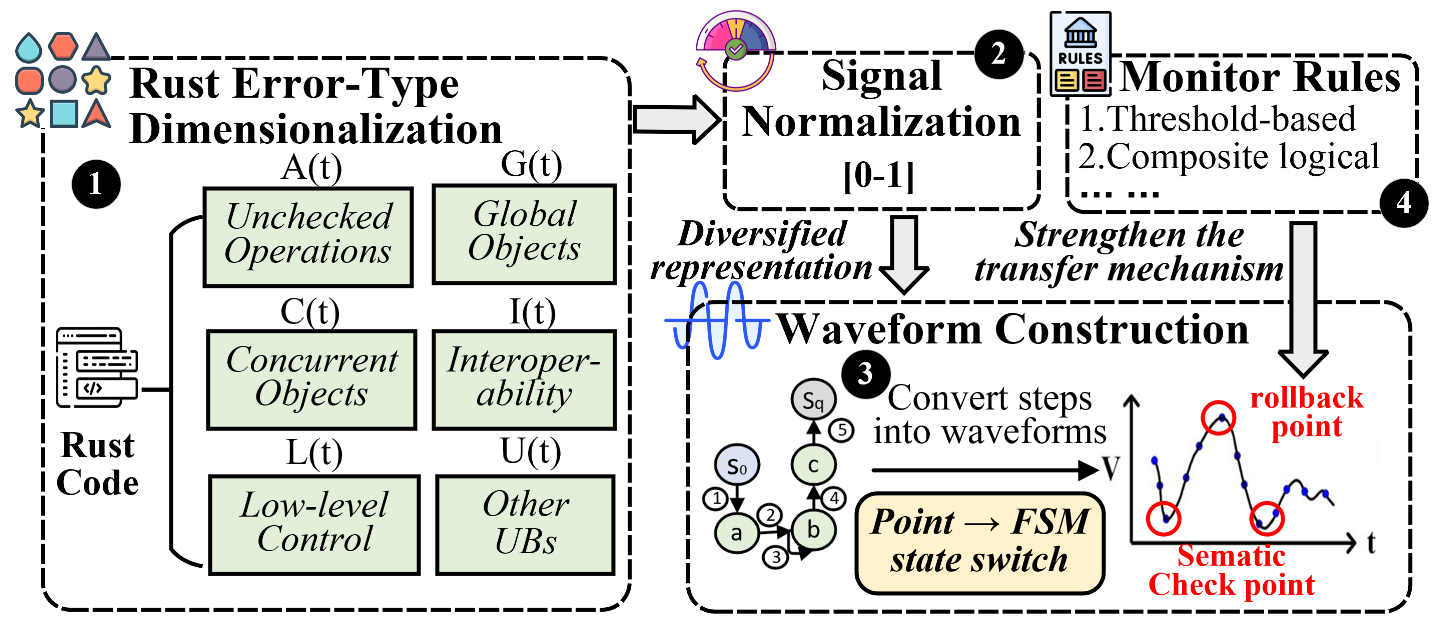}}
\vspace{-1.4em}
\caption{Self-Adaptive Correction Mechanism.}
\vspace{-1.2em}
\label{hallucination}
\end{figure}

Specifically, AkiraRust first collects the runtime context generated after each state transition, including UB reports, semantic, and structural information as logs, and maps them to a fixed-dimensional error-signal vector.  
Based on prior studies~\cite{jiang2025thetis, dong2024safenet,gaher2024refinedrust,jung2021safe}, AkiraRust classifies various types of unsafe code into several typical classifications (\ding{182} in Fig.~\ref{hallucination}), and constructs a fixed-dimensional signal channel set: $f(t)=[U(t),G(t),I(t),L(t),C(t)]$ , where each component corresponds to a specific category of runtime errors, unchecked operations $(U)$, global objects $(G)$, interoperability $(I)$, low-level control $(L)$, and concurrency objects $(C)$. 
Then, by normalizing thees signals $f(t)$ (\ding{183} in Fig.~\ref{hallucination}), AkiraRust evaluates code repair scores, and constructs a unified defect-evolution waveform E(t) over time (\ding{184} in Fig.~\ref{hallucination}). This waveform is a curve of code-incorrectness scores that reflects how repair quality evolves: 
$E(t) = \sum_{i} {w_i} \cdot {g_i}(f_{i}(t))$ . 
Here, $w_i$ denotes each error type’s weight, derived from historical repair frequency and severity, and $g_i(\cdot)$ is a temporal smoothing function capturing error persistence over time. 
Finally, based on the code-incorrectness scores, we define a set of monitoring rules and critical-point criteria to enhancing the state transition mechanism, from which two monitoring points are derived (\ding{185} in Fig.~\ref{hallucination}):  
\textbf{(i) Rollback Points}, triggered when the waveform sharply fluctuates or exceeds a threshold, indicating instability or semantic drift. 
When detected, AkiraRust stops further modifications and rollback to the initial code state. 
\textbf{(ii) Semantic Evaluation Points}, activated when the waveform stabilizes within a low-variance region, AkiraRust  invokes \textit{TestGenAgent} to evaluate the current code’s semantic correctness. 
When these two monitoring points are detected, AkiraRust evaluates the current state semantic or rollback to the lowest-incorrectness score. 

Overall, this design enhances the FSM's transition mechanism in AkiraRust by lightweight monitoring repair dynamics, avoiding redundant semantic checks and unnecessary full rollbacks, reducing runtime overhead while maintaining repair accuracy (Sec.~\ref{dynamic_evaluation}). 

\section{Evaluation} 
\label{dynamic_evaluation}
The evaluation of AkiraRust designed to answer the following research questions: 
\begin{enumerate}
    \item \textbf{RQ1 (Accuracy):} What extent is AkiraRust successful in fixing Rust UBs as measured by metrics of \textit{pass} and \textit{exec}?  
    \item \textbf{RQ2 (Adaptive):} How does AkiraRust's adaptation affect repair semantic correctness? 
    \item \textbf{RQ3 (Advancement):} 
    How does the efficiency and performance of AkiraRust in eliminating UBs compared to the SOTA method and human experts? 
\end{enumerate}

Firstly, we give a brief explanation of the datasets, large language models, baselines and metrics used in this paper. 

\noindent
\textbf{Datasets.}
We use Rust UBs dataset from the Miri compiler~\cite{Miri_dataset}. 
The dataset covers 47 common UB types, including allocation errors, concurrency, data races, and borrow violations, spanning diverse scenarios such as multi-threading and cross-module conflicts.

\noindent
\textbf{LLMs.} 
We use three LLMs: GPT-4~\cite{GPT4}, GPT-5~\cite{GPT5}, and Claude-3.5~\cite{Claude3.5}, each with a default temperature of 0.5.

\noindent
\textbf{Baselines.} 
We select both template-based frameworks (Thetis-lathe~\cite{jiang2025thetis}, TECS'25) and LLM-based Rust repair frameworks (e.g., RustAssistant~\cite{deligiannis2025rustassistant}, ICSE'25), which represent a single-process tool, while RustBrain~\cite{jiang2025unlocking} (DAC'25) serves as a multi-process tool.


\noindent
\textbf{Metrics.} 
All experiments report \textit{pass} rate(\%) and \textit{exec} rate(\%), among which \textit{pass} rate refers to the percentage of code passing Miri, and \textit{exec} rate refers to the percentage of code that passes semantic correctness checks during evaluation by \textit{TestGenAgent}.

\begin{figure}[t]
\centering
\includegraphics[width=\linewidth]{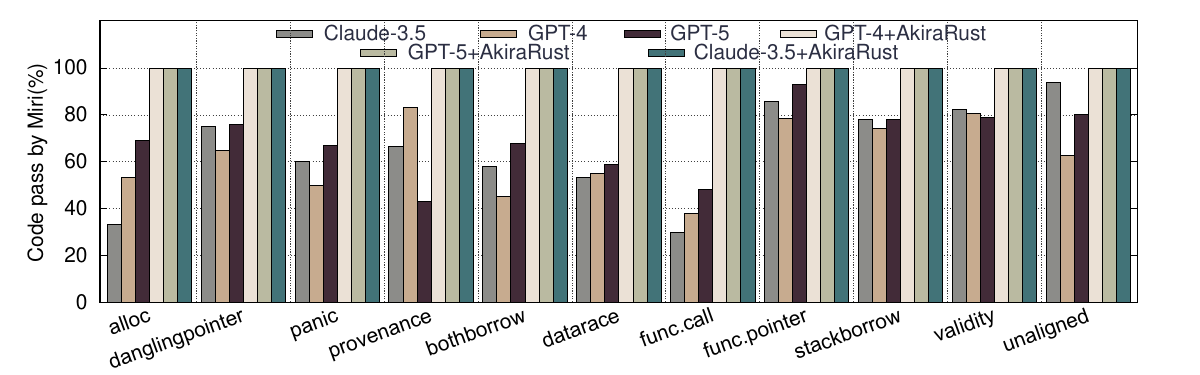}
\vspace{-2.4em}
\caption{AkiraRust Repairs UBs Pass Rate by Miri.}
\vspace{-1.3em}
\label{fig:pass}
\end{figure}

\begin{figure}[t]
\centering
\includegraphics[width=\linewidth]{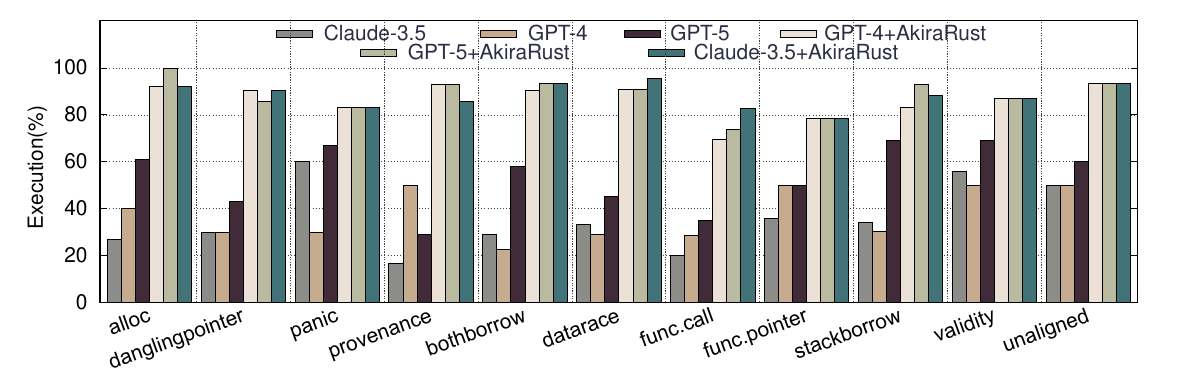}
\vspace{-2.4em}
\caption{AkiraRust Repairs UBs Semantic Execution Rate.}
\vspace{-1.3em}
\label{fig:exec}
\end{figure}

Subsequently, we analyze the results to answer these questions.
\noindent
\textbf{RQ1 (Accuracy):} 
We evaluated AkiraRust's repair \textit{pass} and \textit{exec} rate using various LLM configurations.  
In Fig.~\ref{fig:pass}, using an LLM alone, even GPT-5, only achieves a 75\% \textit{pass} rate, failing to handle complex ownership, lifetime, and aliasing issues in dataset. 
AkiraRust achieves a 100\% \textit{pass} rate, demonstrating that all generated patches compile and satisfy basic functional constraints. 
We further validate semantic correctness.
In Fig.~\ref{fig:exec}, due to Rust’s strict compiler rules, and the lack of sufficient domain-specific training data, directly using general-purpose LLMs (GPT-4, GPT-5, Claude 3.5) for repair achieves only modest semantic \textit{exec} rates (typically 25–50\%). 
In contrast, AkiraRust, across different underlying LLMs, maintains consistently high \textit{exec} rates ($\approx$ 85–95\%), representing a 40–60\% improvement over standalone LLM-based repair approaches.  
We observe that although stronger LLMs (AkiraRust+GPT-5, compared with AkiraRust+GPT-4/Claude-3.5) yield slightly higher \textit{exec} rates, AkiraRust leverages FSM structure and waveform-guided adaptive mechanism to select more suitable reasoning trajectories, narrowing the performance gap between models. 
This makes to uniformly high semantic repair accuracy and effectively decouples repair quality from raw LLM reasoning capacity.  

\noindent
\textbf{RQ2 (Adaptive):}  
AkiraRust’s adaptive are primarily reflected in three aspects: agent selection, LLM reasoning mode, and the occurrence of hallucinated code (Sec.~\ref{motivation}). 
While the first two aspects are validated by repair accuracy, we focus on the self-adaptive correction mechanism. 
Here, we evaluate AkiraRust’s adaptability to different UB types and its ability to mitigate hallucination-driven deviations during the repair process. 

\begin{figure}[t] 
\centerline{\includegraphics[width=0.47\textwidth, trim=0cm 0.1cm 0cm 0cm, clip]{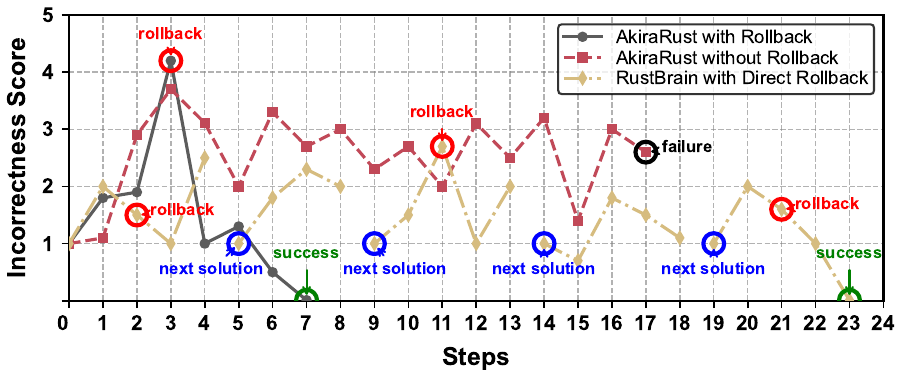}}
\vspace{-0.7em}
\caption{Comparison of AkiraRust Adaptive Rollback, Non-Rollback, and RustBrain Direct Rollback in Repairing Rust UBs.}
\vspace{-1.5em}
\label{dynamic} 
\end{figure}

\begin{figure}[t]
  \centering
  \begin{minipage}[t]{0.23\textwidth}
    \centering
    \footnotesize
    \includegraphics[width=\linewidth]{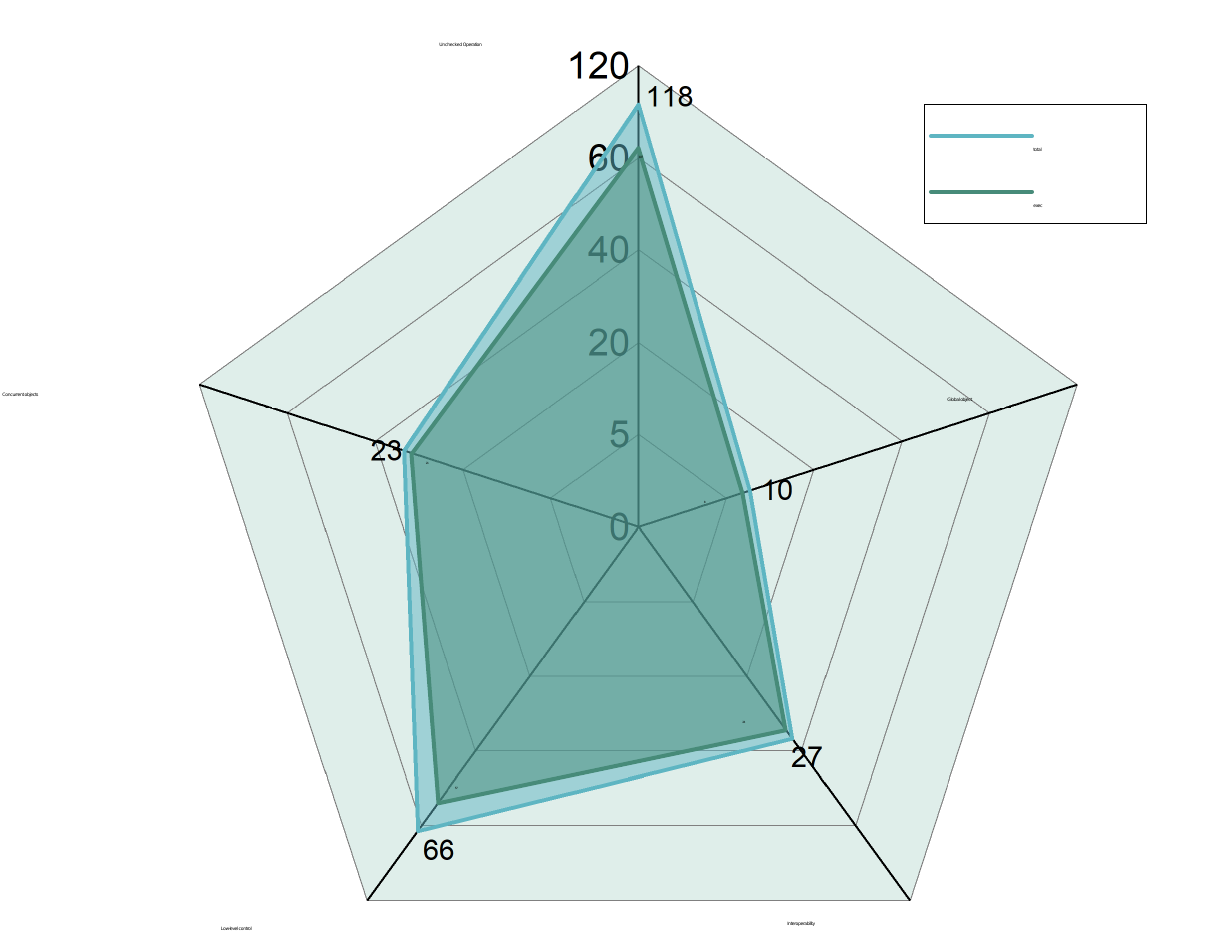}
    \vspace{-2.2em}
    \caption{AkiraRust Repair performance across different Types.}
    \vspace{-1em}
    \label{adaptive1}
  \end{minipage}
  \hfill
  \begin{minipage}[t]{0.24\textwidth}
    \centering
    \footnotesize
    \includegraphics[width=\linewidth]{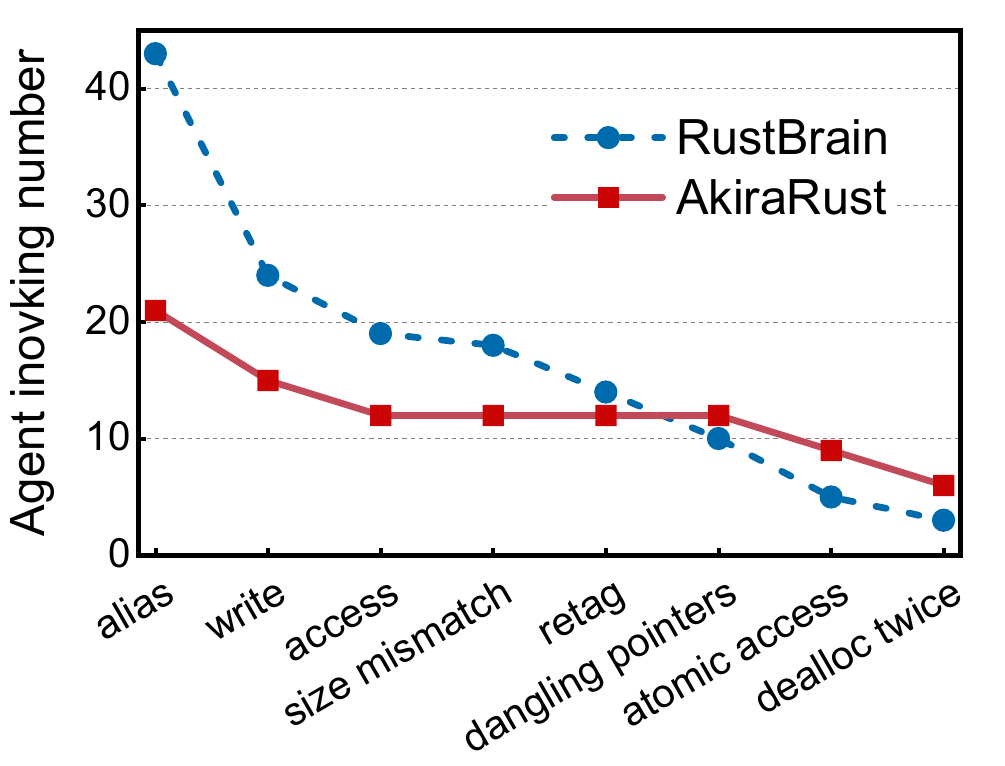}
    \vspace{-2.2em}
    \caption{Comparison of Agent Invocation Counts.}
     \vspace{-1em}
    \label{adaptive2}
  \end{minipage}
  \hfill
\end{figure}

First, we categorize the dataset samples by the Unsafe Rust error types $f(t)$ (See Sec.~\ref{adptive_rollback}) and evaluate AkiraRust’s repair performance within each types. 
In Fig.~\ref{adaptive1}, AkiraRust maintains consistently high and close repair quality across all categories (about 86\%-92\%), indicating that its adaptive mechanism adjusts the reasoning trajectory according to both the error type and context, rather than relying on error-specific. 
Then, we select a representative complex UB case and evaluate it under AkiraRust and RustBrain across three strategies: adaptive rollback (AkiraRust), non-rollback (AkiraRust), and direct rollback (RustBrain). 
In Fig.~\ref{dynamic}, for RustBrain, N solution processes are generated to perform repairs, and a direct rollback occurs whenever a round fails. 
In this case, it takes four rounds (invoking 23 agents) to complete the repair. Without adaptive rollback, AkiraRust would allow hallucinated code to persist unchecked, continuing to propagate until the maximum iteration limit is reached, causing repair failure. 
In contrast, AkiraRust’s adaptive mechanism identifies rollback points to select the most promising repair path, avoiding hallucination spread and completing the repair with only 7 agents invocations. 
Overall, this adaptive design maintains stable repair quality across various UB types, while reducing redundant reasoning and preventing the accumulation of illusions. 

\begin{figure}[t] 
\centerline{\includegraphics[width=0.5\textwidth]{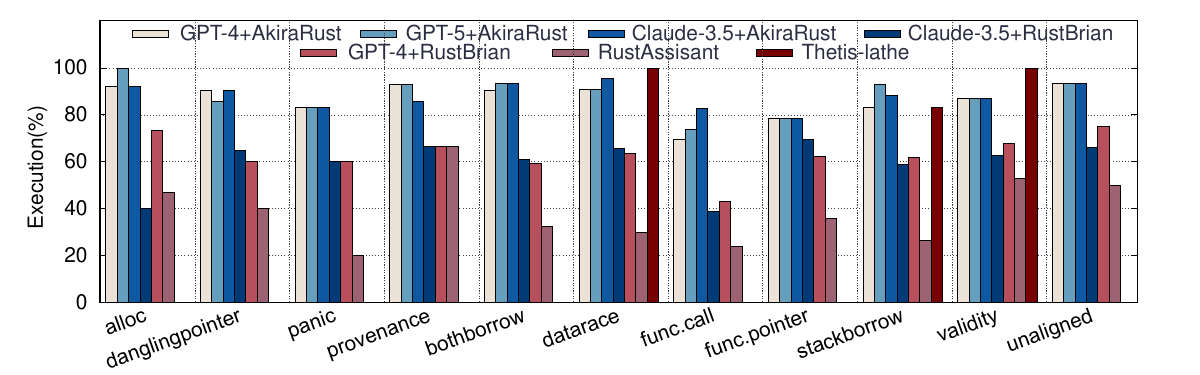}}
\vspace{-1.4em}
\caption{Comparative of AkiraRust and Existing Tools.} 
\vspace{-1.0em}
\label{advancement} 
\end{figure}

\noindent
\textbf{RQ3 (Advancement):} 
To evaluate the effectiveness and efficiency of AkiraRust, we compare it against a static template-based method (Thetis-Lathe~\cite{jiang2025thetis}), LLM-aided methods (RustAssistant~\cite{deligiannis2025rustassistant} and RustBrain~\cite{jiang2025unlocking}, SOTA Rust repair tool). 
In Fig.~\ref{advancement}, AkiraRust also achieves a semantic correctness of 92\%, compared to RustAssistant and RustBrain, the \textit{exec} rate of AkiraRust increased by 15\% and 50\%, respectively, while maintaining performance comparable to Thetis-Lathe. 
However, Thetis-Lathe is restricted to template UB patterns and fails to generalize beyond predefined rules, limiting its applicability to real-world unsafe Rust defects.  

To evaluate AkiraRust's effectiveness, we report the number of agent invocations (Fig.~\ref{adaptive2}) and the execution time (Tab.~\ref{human}), reflecting both repair process cost and runtime efficiency. 
AkiraRust reduces overhead by 7× compared to human experts and shortens repair time by about 2× relative to RustBrain (Fig.~\ref{adaptive2}). 
This is because, for complex errors (e.g., aliasing violations), RustBrain must enumerate and execute multiple solutions, resulting in about 2× the number of agent invocations compared to AkiraRust. 
For simpler repairs (e.g., double dealloc), AkiraRust, due to its lightweight monitoring (Sec.~\ref{hallucination}), invokes agents slightly more frequently than RustBrain, but the additional cost is minimal (green in Tab.~\ref{human}). 
Furthermore, we compare AkiraRust under different LLMs, stronger models (GPT-5) converge faster, yet all variants still outperform RustBrain.

Overall, AkiraRust provides a dynamically adaptive framework to repair Rust UBs, improving semantic repair correctness, adapting its reasoning trajectory to diverse UB types vian FSM-driven agent/mode selection, and reducing repair time and verification overhead compared with SOTA methods and human experts. 



\begin{table}[t]
    \renewcommand{\arraystretch}{1.1}
    \caption{Execution Time of AkiraRust VS. Human and SOTA. (Speepup: RustBrain Time (s)/ AkiraRust+GPT-5 Time (s))}
    \vspace{-1.0em}
    \label{human}
    \centering
    \resizebox{0.95\linewidth}{!}{%
    \begin{tabular}{|c|c|c|c|c|c|c|}
        \hline
        \multirow{3}{*}{\textbf{Types}} & \multicolumn{5}{c|}{\textbf{Average Time/s}} &\multirow{3}{*}{\textbf{Speedup}}\\ 
        \cline{2-6} 
         &  \textbf{Rust} &  \multicolumn{3}{c|}{\textbf{AkiraRust}}   &  \multirow{2}{*}{\textbf{Human}}    &    \\ 
         \cline{3-5} 
         & \textbf{Brain} & GPT-5 & GPT-4 &Claude 3.5& & \\
        \hline
        stack borrow & 254 & 190.6& 149 & 107.5 & 366 &1.3x    \\  
        unaligned & 207 & 90 & 138 & 94.5 & 222  & 2.3x  \\ 
         validity & 255 &  52 &102 & 103 & 678 & 4.9x   \\  
          \rowcolor{teal!26}
         \textbf{alloc} & \textbf{40} & \textbf{63} & \textbf{68} & \textbf{157} & \textbf{450} & \textbf{0.5x} \\ 
         func\_pointer & 160 & 71 & 64 & 207 &480 & 2.3x   \\
         provenance & 183 & 107 & 108 & 150 & 240  & 1.71x  \\  
         panic & 309 & 89 & 182 & 145 & 336   & 3.5x \\  
         \rowcolor{teal!26}
         \textbf{func\_call}& \textbf{151} & \textbf{151} & \textbf{86} & \textbf{161} & \textbf{1176} & \textbf{1x}\\ 
         dangling & 85 & 65 & 50 & 106 & 114 & 1.3x \\ 
         borrow & 243 & 130 & 135 & 220 &762 & 1.9x   \\ 
         data\_race & 294 & 147 & 231 & 173 & 336  &  2.0x \\ \hline
         \rowcolor{gray!17} 
         \textbf{Average} & \textbf{218.4} & \textbf{101.4} &  \textbf{114.6}&\textbf{147.3}& \textbf{442}&\textbf{2.2x}\\
        \hline
    \end{tabular}}
\vspace{-1.0em}
\end{table}

\section{Conclusion}
We've introduced AkiraRust, a runtime and adaptive detection-and-repair tool designed to eliminate UBs and improve repair correctness and semantic accuracy.
Our approach involved leveraging LLMs guided by FSMs to dynamically adapt detection and repair strategies based on runtime semantic context. 
To bring this framework to life, we developed the dual-mode multi-agent system, enabling fast and slow reasoning modes to coordinate multiple agents performing diverse tasks, and designed a state transition mechanism to dynamically adjust agent strategies and task sequences, ensuring semantic consistency and stability throughout the repair process. 
Experimentally, AkiraRust exhibited a significant improvement in UB detection and repair accuracy with less overhead. 

\section{Acknowledgment}
We appreciate the anonymous reviewers for their helpful feedback. 
This paper was written during a special moment in my life. 
Renshuang Jiang sincerely thanks our sweetest little cat, Akira Jiang, for joining our lives and filling them with love. 
Additionally, Shuang would extend her thanks to Zhe Jiang for his continuous support and companionship throughout the years. 
\bibliographystyle{ACM-Reference-Format}
\bibliography{software}

\end{document}